\documentstyle[12pt]{article}
\normalsize

\def\a{\alpha}
\def\b{\beta}

\def\d{\delta}
\def\e{\epsilon}

\def\g{\gamma}

\def\l{\lambda}
\def\m{\mu}
\def\n{\nu}

\def\s{\sigma}

\def\P{\Pi}

\def\S{\Sigma}

\def\bar#1{\overline{#1}}

\def\Hat#1{\rlap{\kern.10em$\widehat{\phantom G}$}#1}
\def\HAt#1{\rlap{\kern.05em$\widehat{\phantom G}$}#1}

\def\cap#1{\rlap{\kern.1em$\widehat{\phantom{G\vrule height.8em}}$}#1{}}
\def\Cap#1{\rlap{\kern.05em$\widehat{\phantom{G\vrule height.8em}}$}#1{}}

\newcounter{sxn}

\newcounter{axn}

\def\br{\begin{thebibliography}{abc}}
\def\er{\end{thebibliography}}
\def\rf{\bibitem}
\date{}

\begin{document}
\bibliographystyle{unsrt}
\footskip 1.0cm
\thispagestyle{empty}
\setcounter{page}{0}
\begin{flushright}
SU-4240-429\\
January 1993\\
\end{flushright}
\vspace{10mm}

\centerline {\LARGE EDGE STATES IN 4d AND THEIR 3d}
\vspace{5mm}
\centerline {\LARGE GROUPS AND FIELDS}
\vspace*{15mm}
\centerline {\large A.P. Balachandran$^{(1)}$, G. Bimonte$^{(1,2)}$
\small and \large
                          P. Teotonio-Sobrinho$^{(1)}$}
\vspace*{5mm}
\centerline {(1)\it Department of Physics, Syracuse University,}
\centerline {\it Syracuse, NY 13244-1130}
\centerline{(2)\it Dipartimento di Scienze Fisiche dell'Universit\`{a}
di Napoli}
\centerline{\it Mostra d'Oltremare pad. 19, 80125 Napoli, Italy}
\vspace*{25mm}
\normalsize
\centerline {\bf Abstract}
\vspace*{5mm}

It is known that the Lagrangian for the edge states of a Chern-Simons
theory describes a coadjoint orbit of a Kac-Moody (KM) group with its
associated Kirillov  symplectic form and group representation.
It can also be obtained from a chiral sector of a nonchiral field theory. We
study the edge states of the abelian $BF$ system in four dimensions (4d) and
show the following results in almost exact analogy: 1) The Lagrangian for
these states is associated with a certain 2d generalization of the KM group.
It describes a coadjoint orbit of this group as a Kirillov symplectic
manifold and
also the corresponding group representation. 2) It can be
obtained from with a
``self-dual" or ``anti-self-dual" sector of a Lagrangian describing a massless
scalar and a Maxwell field [ the phrase ``self-dual" here being used
essentially in its sense in monopole theory]. There are similar results for
the nonabelian $BF$ system as well. These shared features of edge states in
3d and 4d suggest that the edge Lagrangians for $BF$ systems are certain
natural generalizations of field theory Lagrangians related to KM groups.

\newpage

\baselineskip=24pt
\setcounter{page}{1}
\newcommand{\be}{\begin{equation}}
\newcommand{\ee}{\end{equation}}

In a previous paper \cite{bala}, we studied the topological action
$$
S_{BF}= \int _{M^3\times {\bf R}^1}B\wedge F \,,
$$
\be
B=B_{\mu \nu}dx^\mu \wedge dx^\nu \,\, ,F=dA\, , A=A_\mu dx^\mu \label{1}
\ee
where the spatial manifold $M^3$ is a manifold with boundary $\partial M^3$,
such as a solid ball ${\cal B}^3$ or a solid torus ${\bf T}^3$. It was
established that there are states localised on $\partial M^3$ which carry a
representation of a certain Lie algebra. If the superscript $(j)$ indicates
$j$ forms, this algebra can be described in terms of the following
commutators in its representation $\rho \,\, : \,\, \lambda ^{(j)}
\rightarrow \rho (\lambda ^{(j)})$ on quantum states:
$$
\left[ \rho (\lambda ^{(0)}),\rho (\mu ^{(0)})\right]=
\left[ \rho (\lambda ^{(1)}),\rho (\mu ^{(1)})\right]= 0\,,
$$
\be
\left[ \rho (\lambda ^{(0)}),\rho (\lambda ^{(1)})\right]=
\imath \int _{\partial M^3}\lambda ^{(0)}d\lambda ^{(1)} .\label{2}
\ee
[In ref. \cite{bala}, $\rho(\lambda ^{(j)})$ were written as integrals
$\int _{M^3}d\lambda ^{(0)}B$, $\int _{M^3}d\lambda ^{(1)}A$ involving $B$ or
$A$ when $d\lambda ^{(j)}$ were nonzero on $\partial M^3$,
$\lambda ^{(j)}$ here being extentions of $\lambda ^{(j)}$ in (\ref{1})
to all of $M^3$.] The subalgebra with generators $\rho (\lambda ^{(0)})$ is
the algebra of the group of maps $\partial M^3\rightarrow U(1)$
[or ${\bf R}^1$] while
$\rho (\lambda ^{(1)})$ are the generators of another abelian subalgebra.
(\ref{2}) describes an extention of the direct sum of these subalgebras by
the abelian Lie algebra of reals.

We now recall that the abelian Chern-Simons (CS) action on $({\rm disk}\,\,
{\bf M}^2)\times {\bf R}^1$ also produces an algebra on the bounding circle
$\partial {\bf M}^2$ of ${\bf M}^2$ \cite{wit}. [See also ref. \cite{bimo} and
\cite{sri} and
references therein.] It is the $U(1)$ KM algebra spanned by functions
$\Lambda $ on $\partial M^2$ (and a central charge $k$). If
$\varphi \,\,:\,\, \Lambda \rightarrow \varphi (\Lambda )$ describes an
irreducible representation of this algebra, it is characterized by the
commutators
\be
\left[\varphi (\Lambda ), \varphi (\tilde \Lambda )\right]=\imath
\frac {k}{2\pi}\int _{\partial M^2}\Lambda d\tilde \Lambda \,,\label{3}
\ee
$k$ having a constant value in the representation. In this note, we show that
there are several features common to (\ref{2}) and (\ref{3}).

The properties of the algebra defined by (\ref{2}) which are of interest here
are the following:

1) It can be produced by canonically quantising the action
\be
\frac {k}{4\pi }\int dt\int _{\partial {\bf M}^2}\partial _t \chi d\chi
 \label{4}
\ee
where $d$ does not involve differentiation in $t$.
[Cf. references \cite{Rai}, \cite{Ale} and citations therein.]

2) An element of the algebra described by (\ref{3}) is a pair $(\Lambda ,\xi )$
, $\xi $ being a real number. An
element of its dual can be written as $(\sigma ^{(1)},\eta )^*$, $\sigma
^{(1)}$
being a one form and $\eta $ a real number, on introducing the pairing
\be
\langle (\sigma ^{(1)},\eta )^*,(\Lambda ,\xi )\rangle = \int _{\partial M^2}
\sigma ^{(1)}\Lambda + \eta \xi \,\,. \label{5}
\ee
Now there is a natural action of the KM group ${\cal K}=\{h\}$ on
$(\sigma ^{(1)},\eta )^*$ called the coadjoint action. If
$(\Lambda ,\xi )\rightarrow h(\Lambda ,\xi )h^{-1}$ is the adjoint action, the
coadjoint action ${\rm Ad}^*~h$ of $h$ is defined by
\be
\langle {\rm Ad}^*\,h(\sigma ^{(1)},\eta )^*,h(\Lambda , \xi )h^{-1}\rangle =
\langle (\sigma ^{(1)},\eta )^*,(\Lambda , \xi )\rangle . \label{6}
\ee
{}From a general result of Kirillov \cite{kir}, it is known that an orbit of
${\cal K}$
for this coadjoint action (a ``coadjoint" orbit) carries a $\cal K$-invariant
symplectic form $\omega ^{(2)}$. For the orbit through $(0,1)^*$, a simple
calculation \cite{Rai,Ale}
also shows that we can write $\omega ^{(2)}=d\omega ^{(1)}$. We
can thus contemplate forming the action
\be
\int \omega ^{(1)} \label{7}
\ee
which is like the action $pdq$ in particle mechanics. [See ref. \cite{mar} and
references therein.] (\ref{7}) can be brought to the form (\ref{4}) up to
surface terms \cite{Rai,Ale}.

3) The current algebra defined by (\ref{3}) can be obtained from the scalar
field Lagrangian
\be
\frac {|k|}{8\pi }\int _{\partial M^2}d\theta [(\partial _t\phi )^2-
(\partial _\theta \phi )^2 ]\label{8}
\ee
by imposing either of the constraints
\be
\partial _{\pm}\phi =0,\,\,\, \partial _\pm =
\partial _t \,\,\pm \,\,\partial _\theta \label{9}
\ee
depending on the sign of $k$.[ Here $\theta $(mod $2\pi $) is the coordinate
on the circle while the speed for the field $\phi $ has been set equal to 1.]
Any solution of the field equation for (\ref{8}) is in fact
the sum $\phi _++\phi _-$ where $\partial _-\phi _+=
\partial _+\phi _-=0$.

4) The Hamiltonian for $(\ref{4})$ is zero whereas that is not the case for
(\ref{8}). The latter in fact evolves $\phi $ preserving the condition in
(\ref{9}). This evolution of left- and right-movers are also given by the
nonlocal Lagrangians \cite{chi}
$$
\frac {\pi }{k}\int _0^{2\pi }d\theta d\theta '\phi (\theta ,t)\epsilon
(\theta -\theta ')
\partial _t\phi (\theta ',t) \pm \frac {2\pi }{k}\int _0^{2\pi }d\theta
\phi (\theta ,t)^2\,,
$$
\be
\epsilon (\theta -\theta ')=-\epsilon (\theta '-\theta )=1\,\,\,{\rm if}\,\,\,
\theta >\theta '\,.\label{10}
\ee

In this note, we will show that each of these features can be generalised to
the algebra defined by (\ref{2}). There are similar generalisations for the
nonabelian
problem as well as we shall later indicate. All this suggests that the 3d
systems
coming from 4d topological actions are certain natural generalisations of 2d
systems associated with KM groups.

For simplicity of
presentation when discussing the generalizations of 1) to 4)
under items 1) to 4) below, it is covenient to assume
that $M^3$ is the solid torus ${\bf T}^3$.
For $\partial M^3=T^2$ (the two torus),we also choose the flat metric
$(d\theta ^1)^2+
(d\theta ^2)^2$ [$\theta ^i$ mod $2\pi $ being the standard coordinates on
$T^2$].

\vspace{10mm}
\noindent
{\large {\bf Item 1}}
\vspace{6mm}

For the current algebra (\ref{2}), the Lagrangian is
\be
\int _{\partial M^3}\left(\partial _t\phi dA+ d\phi
\partial _t A \right)\,\,,\label{11}
\ee
where
\be
A=A_jd\theta ^j, \,\, d\phi= \partial _j\phi d\theta ^j, \,\,
dA=\partial _kA_jd\theta ^k\wedge d\theta ^j \equiv
\partial _k A_j d\theta ^kd\theta ^j\,. \label{12}
\ee
[Wedge symbols between differential forms will hereafter be omitted.]
Note that as before $d$ does not differentiate time $t$. Also we can
equally well consider the negative of the Lagrangian (\ref{11}).

This result can be shown as follows. If $\Pi$ and $P^j$ are the momenta
conjugate to $\phi $ and $A_j$, (\ref{11}) leads to the constraints
\be
\Pi - *dA := \Pi - \epsilon ^{ij}\partial _iA_j\approx 0\,\,,
{}~~~~~~~~~~
P_i+(*d\phi )_i:= P_i+\epsilon _{ij}\partial ^j\phi
\approx 0  \,\,,\label{13}
\ee
where $\epsilon _{ij}=-\epsilon _{ji},\,\, \epsilon _{12}=1$, the spatial
metric is $(1,1)_{\rm diagonal}$ and
$\approx $ denotes weak equality. The first class variables or
observables with zero Poisson brackets (PB's) with these constraints are
functions of
\be
\Pi + *dA\,,  \:\:\:\:\:P_i -(*d\varphi )_i   \,.\label{14}
\ee
We can now set
$$
\rho(\lambda ^{(0)})=\int _{T^2}\lambda ^{(0)}(*\Pi + dA)\,,
\:\:\:\: \rho (\lambda ^{(1)})=-\frac 12\int _{{\rm T}^2}\lambda ^{(1)}
(*P+ d\phi )\,,
$$
\be
*\Pi:=\Pi d^2\theta \,, \:\:\:\: *P:=\epsilon _{ij}d\theta ^iP^j\,, \label{15}
\ee
as they have the commutators (\ref{2}) in quantum theory.

\vspace{10mm}
\noindent
{\large {\bf Item 2}}
\vspace{6mm}

We first outline the general method to construct the Kirillov symplectic
form and its associated one form. Let $G=\{g\}$ be a Lie group with Lie
algebra $\underline{G}=\{\alpha \}$. If $\underline{G}^*=\{\beta ^*\}$ is
the dual
of $\underline{G}$, we denote the pairing of $\beta ^*$ and $\alpha $ by
$<\beta ^*,\alpha >$. If $\alpha \rightarrow g\alpha g^{-1}$ is the adjoint
action of $g$, its coadjoint action ${Ad}^*\,g$
is defined analogously to (\ref{6}) by
requiring
\be
\langle {\rm Ad}^*\,g\,\beta ^*,g\alpha g^{-1}\rangle =
\langle \beta ^*,\alpha \rangle . \label{16}
\ee

With this action, $G$ defines orbits (``coadjoint orbits") in
$\underline{G}^*$. The coadjoint orbit through $K^*$ with stability group $H$
can be identified with the coset space $G/H$ in a well-known way.

Now consider the one form
\be
\Omega ^{(1)}=\imath <K^*,g^{-1}dg> \label{17}
\ee
on $G$.
Then as frequently explained elsewhere \cite{mar}, the associated two form
$d\Omega ^{(1)}$ projects down to a two form $\Omega ^{(2)}$ on $G/H$ and that
form is the Kirillov symplectic form on $G/H$.
Furthermore, according
to our previous work \cite{mar}, and analogously to (\ref{7}), the Lagrangian
leading to this form is
\be
\imath <K^*,g^{-1}\partial_tg>. \label{18}
\ee
We now show that (\ref{18}) is exactly (\ref{11}) for a suitable choice of
$K^*$.

A general element of the Lie algebra $\underline {\cal G}$ for (\ref{2})
can be written as $(\lambda ^{(0)},\lambda ^{(1)},\xi )$
[$\xi \in R^1$]. The Lie bracket is given by
\be
\left[ (\lambda ^{(0)},\lambda ^{(1)},\xi ),(\mu ^{(0)},\mu ^{(1)},\nu )
\right]= (0,0,\imath \int \lambda ^{(0)} d\mu ^{(1)}
-\imath \int \mu ^{(0)} d\lambda ^{(1)})\,. \label{19}
\ee
A general element of the group $\cal G$ with Lie algebra $\underline{\cal G}$
is $g(\lambda ^{(0)},\lambda ^{(1)},\xi )=exp(\imath \lambda ^{(0)})
exp(\imath \lambda ^{(1)})exp(\imath \xi )$, $exp$ being the usual exponential
map, and the indicated order of factors in writing
$g(\lambda ^{(0)},\lambda ^{(1)},\xi )$ will hereafter be adopted as a
convention. The adjoint group action can be worked out using (\ref{19}):
\be
g(\lambda ^{(0)},\lambda ^{(1)},\xi )(\mu ^{(0)},\mu ^{(1)},\nu )
g(\lambda ^{(0)},\lambda ^{(1)},\xi )^{-1}= (\mu ^{(0)},\mu ^{(1)},\nu +
\int d\lambda ^{(0)}\mu ^{(1)}+\int d\lambda ^{(1)}\mu ^{(0)}). \label{20}
\ee

Let $\underline{\cal G}^*$ be the dual of $\underline{\cal G}$. Its elements
can be written as $(\s^{(2)},\s^{(1)},\eta)^*$, $\s^{(j)}$ being $j$ forms
and $\eta \in R^1$. The pairing between elements of $\underline{\cal G}^*$
and $\underline{\cal G}$ here is
\be
\langle (\s^{(2)},\s^{(1)},\eta)^*,(\m^{(0)},\m^{(1)},\nu )\rangle=
\int_{\partial M^3} (\s^{(2)} \mu ^{(0)}+\s^{(1)}\mu ^{(1)})+\eta \nu ).
\label{21}
\ee

The coadjoint action now follows from (\ref{16}):
\be
{\rm Ad}^*\,g\,(\l^{(0)},\l^{(1)},\xi)(\s^{(2)},\s^{(1)},\eta)^*=
(\sigma ^{(2)}-\eta d\lambda ^{(1)},\sigma ^{(1)}-\eta d\lambda ^{(0)},\eta ).
\label{22}
\ee

With the choice $K^*=(0,0,2)$, the Lagrangian (\ref{11}) readily follows from
(\ref{22})and (\ref{18}) [with $\phi =\lambda ^{(0)}$, $A=\lambda ^{(1)}$ ]on
noting that
\be
g(\l^{(0)},\l^{(1)},\xi)^{-1}\partial _tg(\l^{(0)},\l^{(1)},\xi)=\imath
(\partial _t\lambda ^{(0)},\partial _t \lambda ^{(1)},\partial _t\xi -
\int \partial _t \lambda ^{(0)}d\lambda ^{(1)}).\label{23}
\ee
and discarding certain total derivatives.

The formulae for $\rho(\l^{(j)})$ in (\ref{15}) involve pairings like in
(\ref{21}),
with $*\Pi+dA$ corresponding to $\s^{(2)}$ and $*P +d\phi$
corresponding to $\s^{(1)}$. Thus $\rho$ can be identified with an
element $[*\Pi+dA,*P+d\phi,c]^* := [\S^{(2)},\S^{(1)},c]^*$ ($c\in {\bf R}^1$)
of the dual of $\underline {\cal G}$ with values $\underline{{\rm in~a
{}~representation}}$
(and not
real numbers as for $\underline{\cal G}^*$), the pairing being
$$
([\S^{(2)},\S^{(1)},c]^*,(\l^{(0)},\l^{(1)},\xi))=
\int_{\partial M^3} (\S^{(2)} \l^{(0)}+\frac 12
\S^{(1)}\l^{(1)})+{\bf 1}c\xi \,,
$$
\be
{\bf 1}={\rm unit~~~operator}\,.\label{24}
\ee
Comparison of (\ref{2}) and (\ref{19}) also shows that $c=1$ for our $\rho $.
It is interesting that the relations (\ref{2}), which mean that $\rho$ is a
representation, can be stated as a closure property of $\rho$ or of
$[\S^{(2)},\S^{(1)},c]^*$ in a certain cohomology \cite{Ma2}. Thus, the quantum
fields are operator valued distributions on $\underline {\cal G}$
with a certain closure property signifying that they lead to a
representation of $\underline {\cal G}$.

\vspace{10mm}
\noindent
{\large {\bf Item 3}}
\vspace{6mm}

Consider the Lagrangian
\be
-\int_{\partial M} [\frac{1}{2} \partial_{\mu}\phi\partial^{\mu}\phi+
\frac{1}{4}F_{\mu \nu}F^{\mu \nu}]~~,~~~F_{\mu \nu}=\partial_{\mu} A_{\nu}-
\partial_{\nu} A_{\mu}\label{25}
\ee
where the metric is $(-1,1,1)_{\rm diagonal}$. It leads to the equations of
motion
\be
\partial_{\mu}\partial^{\mu}\phi=0,~~~~~~~~\partial_{\mu}F^{\mu \nu}=0.
\label{26}
\ee
The momenta $\Pi$ and $P^i$ conjugate to $\phi$ and $A_i$ for (\ref{25}) are
given by
\be
\Pi=\partial_0 \phi,~~~~~~P_i=F_{0i},\label{27}
\ee
$P_i$ being subject to the first class constraint
\be
\partial_i P^i \approx 0.\label{28}
\ee

Analogous to the chiral constraints (\ref{9}), we now consider the so-called
``self-dual" or ``anti-self-dual" constraint
\be
\partial_{\mu} \phi= \pm \frac 12\e_{\mu \nu \l} F^{\n \l}\label{29}
\ee
similar to the Bogomol'nyi-Prasad-Sommerfield equations \cite{mono},
where we adopt the convention $\epsilon _{012}=+1$ for the Levi-Cevita symbol.

It is readily seen that (\ref{29}) implies (\ref{26}). If we now rewrite
(\ref{29}) for the
plus sign in terms of $\P$ and $P_i$, we get exactly (\ref{13}). As for the
minus
sign in (\ref{29}), it corresponds in a similar way to the negative of the
Lagrangian (\ref{11}), the equations replacing (\ref{13}) for the latter being
\be
\P + *dA \approx 0,~~~~P_i-\e_{ij}\partial^j \phi \approx 0.\label{30}
\ee

We next show that any solution of (\ref{26}) is the sum of two pieces
$(\phi^{(\pm)},\Pi^{(\pm)},*dA^{(\pm)},P^{(\pm)}:=d\theta ^iP_i^{(\pm)})$,
the fields with the plus
(minus) sign fulfilling (\ref{13})[(\ref{30})] for the
plus (minus) sign. For this purpose, first note that the field equations
(\ref{26}) and the identifications (\ref{27}) give the
following equations:
$$
\partial_0[\P \mp *dA]=\mp \partial_i \e_{ij}
[P_j \pm \e_{jk}\partial^k \phi]\,,
$$
\be
\partial_0[P_i \pm \e_{ij}\partial^j \phi]=\pm \epsilon _{ij}\partial_j
[\P \mp *dA]\,.\label{31}
\ee
They show that the field equations (\ref{26}) preserve the constraints
(\ref{13}) and (\ref{30})
during time evolution without generating new constraints. Hence, it is
sufficient to show that the (gauge invariant) initial data $(\phi, \P,
*dA,P:=d\theta ^iP_i)$ at a fixed time $t_0$ can be written as the sum of two
pieces
$(\phi^{(\pm)}, \P^{(\pm)},*dA^{(\pm)},P^{(\pm)})$, the fields with the plus
(minus) signs fulfilling (\ref{13}) and (\ref{30}).

Now, in view of (\ref{28}), we can write
\be
P_i=\e_{ij}\partial^j f^{(0)}~~~{\rm or}~~P=d\theta ^i\epsilon _{ij}
\partial^jf^{(0)}:=*df^{(0)}~~~{\rm at}~ t=t_0 \label{32}
\ee
for some function $f^{(0)}$.

Let us next consider the initial data
\be
(\phi^{(+)},\P^{(+)},*dA^{(+)}=\P^{(+)},P^{(+)}=-*d\phi^{(+)})+
(\phi^{(-)},\P^{(-)},*dA^{(-)}=-\P^{(-)},P^{(-)}=*d\phi^{(-)})
\label{33}
\ee
at time $t_0$. The first bracket of fields fulfills (\ref{13}) and the second
bracket (\ref{30}). In order that (\ref{33}) equals $(\phi, \P,*dA,P)$ at
$t_0$, it is enough to choose the fields in (\ref{33}) to satisfy
$$
\phi^{(+)}+\phi^{(-)}=\phi,~~~~~\P^{(+)}+\P^{(-)}=\P,
$$
\be
\phi^{(+)}-\phi^{(-)}=-f^{(0)},~~~~~~\P^{(+)}-\P^{(-)}=*dA\label{34}
\ee
at time $t_0$.

We thus see that any solution of (\ref{26}) is the sum of solutions
satisfying (\ref{13}) and (\ref{30}).

The preceding analysis is ``local" and does not address issues that
arise from possible global observables such as Wilson loop integrals and
the possible nonexactness of the closed form $*P$.

\vspace{10mm}
\noindent
{\large {\bf Item 4}}
\vspace{6mm}

The three dimensional analogue of the Lagrangians (\ref{10}) is local and
is the first order form
$$
\int d^3x~{\cal L}\,,
$$
\be
{\cal L} = \epsilon ^{\mu \nu \lambda }B_\mu \partial _\nu
A_\lambda -\frac 12 B_\mu B^\mu  \label{35}
\ee
of the Maxwell Lagrangian.
This is because
the equations of motion for (\ref{35}) are
\be
B_\mu =\frac 12\epsilon _{\mu \nu \lambda}F^{\nu \lambda }\,,\label{36-a}
\ee
\be
\epsilon ^{\mu \nu \lambda }\partial _\nu B_\lambda =0 \,.\label{36}
\ee
As under item 3) for $*P$, we now assume that $B_\mu =\pm \partial _\mu \phi $
for some $\phi $ in view of (\ref{36}). Then (\ref{36-a})
becomes equivalent to (\ref{29}).

It is interesting that (\ref{36-a}) and (\ref{36}) together can be written as a
$2+1$ dimensional
variant of the Duffin-Kemmer-Petiau  equation \cite{duffin}. For this purpose,
we define
$$
\Psi= \left( \begin{array}{c} A_0 \\ A_1 \\ A_2 \\ B_0 \\ B_1 \\ B_2
             \end{array}
      \right) \,, \;\;\;\;\;
\alpha = \left( \begin{array}{cc} 0 & {\bf 1} \\
                                  0 & 0
                     \end{array}  \right)\,,
$$
$$
\beta _\mu= \left( \begin{array}{cc} \tilde \beta _\mu & 0 \\
                                     0 &  \tilde \beta _\mu
                  \end{array}  \right) \,, \;\;\;\;
{\left( \tilde \beta _\mu \right)_\nu }^\lambda ={\epsilon _{\mu \nu }}
^\lambda \,;
$$
\be
\beta _\mu ^{\dag }=\eta _{\mu \mu}\beta _\mu\,\,{\rm (no}~\mu~{\rm sum)},
\label{37}
\ee
where ${\bf 1}$ is the $3\times 3$ unit matrix.
The $\beta $ matrices satisfy the Duffin-Kemmer-Petiau algebra since the
$\tilde\beta $ do so:
\be
\tilde \beta _\mu \tilde \beta _\nu \tilde \beta _\lambda +
\tilde \beta _\lambda \tilde \beta _\nu \tilde \beta _\mu = \eta _{\mu \nu }
\tilde \beta _\lambda + \eta _{\lambda \nu }\tilde \beta _\mu \,.\label{38}
\ee
With (\ref{37}),(\ref{36-a}) and (\ref{36}) are
\be
\left( \beta _\mu \partial ^\mu + \alpha \right)\Psi =0\,. \label{39}
\ee
The Lagrangian density in (\ref{35}) can now also be written in the Dirac form
$$
{\cal L}=-\frac 12\bar{\Psi }(\beta _\mu \partial ^\mu +
\alpha )\Psi \,,
$$
\be
\bar{\Psi }=\Psi ^{\dag }\gamma \,, \:\:\:
\Psi ^{\dag }=(A^0, A^1, A^2, B^0, B^1, B^2)\,, \:\:\:
\gamma = \left( \begin{array}{cc} 0 & {\bf 1} \\
                                 {\bf 1} & 0
                \end{array}  \right) \label{40}
\ee
after discarding a surface term. Here $\Psi$ of course is real.

There are certain generalizations of these considerations to the nonabelian
case. They will only be sketched here, as we plan a more thorough treatment
of the nonabelian problem elsewhere. Let $G$ be a simple compact Lie group
thought of concretely as a group of unitary matrices. Let $\underline G$
be its Lie algebra with basis $\{T(\a)|~\a=1,2 \dots,{\rm dimension}~[G]~
{\rm of}~G\}$ which fulfills $[T(\a),T(\b)]=ic^{\g}_{\a
\b}T(\g),~~TrT(\a)T(\b)=
N \d_{\a \b}~~ (N={\rm constant})$ and $T(\a)^{\dagger}=T(\a)$. In the
nonabelian generalization of (2), $\l^{(j)}, \mu^{(j)}$ become $\underline G$
valued $j$ forms on $\partial M^3$ with $\l^{(j)}=\imath \l_{\a}^{(j)}T(\a)$,
$\l^{(j)}_\a $
being real valued forms, while the commutators in (2) become
$$
[\rho (\l^{(0)}),\rho (\mu^{(0)})]=\rho ([\l^{(0)},\mu^{(0)}])\,,
$$
$$
[\rho (\l^{(0)}),\rho (\l^{(1)})]=\rho ([\l^{(0)},\l^{(1)}])+
\imath \int_{\partial M^3 }
Tr \l^{(0)}d \l^{(1)}\,,
$$
\be
[\rho (\l^{(1)}),\rho (\mu^{(1)})]=0\,.\label{41}
\ee
This gives the generalization of (19) as well. The generalization of (11)
involves a field $u$ valued in $G$ and a one form $W$ valued in $\underline G$
and reads up to an overall constant,
\be
\imath Tr \int_{\partial M^3}\left[  u \partial_t u^{-1}( dW +udu^{-1}W+W
udu^{-1})+ udu^{-1} \partial_t W\right]\label{42}
\ee
while (13,15) must be replaced by their natural nonabelian versions. The method
using coadjoint orbits also leads to (\ref{42}). The Lagrangians replacing
(\ref{25}) and (\ref{35}) are more involved and will be reported elsewhere.

This work was supported by the U.S. Department of Energy under Contract
Number DE-FG02-85ER40231. G.B. had further support from the
Dipartimento di Scienze Fisiche dell'Universit\`{a}
di Napoli. P.T.S. also thanks CAPES (Brazil) for
partial support.

\end{document}